\newcommand{\beq}{\begin{eqnarray}}
\newcommand{\eeq}{\end{eqnarray}}

\newcommand{\eps}{\epsilon}
\newcommand{\half}{{1\over 2}}
\newcommand{\e}{{\rm e}}

\newcommand{\del}{\partial}
\renewcommand{\d}{\partial}

\renewcommand{\theequation}{\thesection.\arabic{equation}}

\newcommand{\cl}{\centerline}

\newcommand{\btem}{\bibitem}

\newcommand{\PL}{Phys.\ Lett.\ {\bf B}}
\newcommand{\PTP}{Prog.\ Theor.\ Phys.}

\newcommand{\PRL}{Phys.\ Rev. \ Lett.}

\documentstyle[12pt]{article}
\begin{document}
\begin{flushright}
RYUTHP-95/4 \\
August, 1995 \\
\end{flushright}

\begin{center}
{\large {\bf  A Geometrical Formulation of
the Renormalization Group Method for Global Analysis II:}}
{\large  Partial Differential Equations}
\end{center}

\vspace{1cm}

{\cl {Teiji Kunihiro}}

\bigskip

\begin{center}

Faculty of Science and Technology, Ryukoku University,\\
Seta, Ohtsu-city, 520-21, Japan\\
\end{center}

\begin{abstract}
It is shown that  the renormalization group (RG) method for global analysis
can be formulated in the context of the classical theory of envelopes:
Several  examples from partial differential equations are analyzed.
The amplitude equations which are usually
derived by the reductive perturbation theory are shown to be naturally derived
as the equations describing the envelopes of the  local solutions obtained in
the perturbation theory.
\end{abstract}

\setcounter{equation}{0}
\section{Introduction}
\renewcommand{\theequation}{\thesection.\arabic{equation}}

Renormalization group (RG) equations\cite{rg} appear in various
 fields of science. In quantum field theory\cite{JZ},
the RG
equation improves
 results obtained in the perturbation theory.  In recent years, the
improvement of the effective potential\cite{JZ,jona,coleman} has acquired
a renewed
interest\cite{kugo}.
The RG equation has also a remarkable success
 in statistical physics especially in the critical phenomena \cite{JZ}.
 One may also note that there is another successful theory of the
critical
 phenomena called coherent anomaly method (CAM)\cite{CAM}; CAM utilizes
 {\em envelopes} of susceptibilities and other thermodynamical
 quantities as a function of temperature. It is well known that
Feigenbaum \cite{Feigenbaum} applied RG equation to deduce a universality of
 some chaotic phenomena.

Recently, Illinois group\cite{goldenfeld1,goldenfeld2} have shown that the
RG equation
 can be also used for non-quantum mechanical problems:  They proposed
 to use the RG equation  to get an asymptotic
 behavior of solutions of differential equations
  including ones of singular and reductive perturbation
 problems in a unified way. Mathematically,  the RG equation is used to
improve the global behavior of the
 local solutions which were obtained in the perturbation theory.
This fact suggests that  the RG method may be  formulated
 in a purely mathematical way  without recourse to
 the notion of the RG. In the previous paper\cite{kunihiro},
we showed that the RG method can be formulated in the context of the
classical theory of envelopes\cite{courant}:
 We pointed out  that
the RG equation is nothing but the envelope equation, and  gave a proof
 why the RG equation can give a globally improved
 solution to ordinary differential equations(ODE's). In fact,
if  a family of the curves $\{C_{\tau}\}_{\tau}$
in the $x$-$y$ plane  is  represented by the
 equation $F(x, y, \tau)=0$,
  the equation $G(x, y)=0$ representing the
envelope $E$ is given  by eliminating the parameter $\tau$ from the  equation
\beq
\frac{\partial F(x, y, \tau)}{\partial \tau}\Biggl\vert _{\tau =x}=0.
\eeq
Here we have chosen the parameter to be the $x$-coordinate of
the point of the tangency of a
 curve $C_{\tau}$ and the envelope $E$.
 The relevance of the envelope equation Eq.(1.1) and the
 (Gell-Mann-Low type) RG equations is evident; the parameter $\tau$
 corresponds to the renormalization point.
Thus one  would also readily recognize that improving the effective
 potential in
 quantum field theory is constructing  the envelope of the effective
potentials  with the renormalization point varied.\cite{kunihiro}

The  purpose of the present paper is to show that the formulation can
be naturally extended to
 a class of partial differential equations (PDE's);
 the PDE's dealt in the present work include
a dissipative nonlinear hyperbolic equation, one- and two-dimensional
 Swift-Hohenberg equations, damped Kuramoto-Shivashinsky
 equation\cite{cross}  and Barlenblatt equation\cite{baren}.
 Most of these examples were examined by
 the Illinois group \cite{goldenfeld1,goldenfeld2,goldenfeld3}.
 Therefore there will be some overlaps
 in the exposition with theirs.
 It is inevitable, however,  because our purpose is to give a purely
mathematical formulation of their method without recourse to the notion of
 the renormalization group.
 We will show  that the amplitude equations usually obtained by the
reductive perturbation methods are  given naturally
 as envelope equations.
In the course of the formulation, relevance of the characteristic manifold
 as the branch stripmanifold will be indicated
to constructing the global solutions.
In Appendix A, we give a short summary of the classical theory of
 envelope surfaces; for PDE's, envelope {\em surfaces} may be more relevant
 in some cases than
 envelope {\em curves} discussed in \cite{kunihiro}.
In Appendix B, we examine Barenblatt's equation in our approach;
 for this equation, only the anomalous exponent of the long-time
 behavior of the solution is given.

\setcounter{equation}{0}
\section{Dissipative nonlinear hyperbolic equation}
\renewcommand{\theequation}{\thesection.\arabic{equation}}

We first consider the following slightly dissipative nonlinear
 hyperbolic equation\cite{goldenfeld2} to apply the envelope
 theory\footnote{See Appendix A and \cite{kunihiro}
for classical theory of envelopes.} to construct a global solution:
\beq
\frac{\del u}{\del t}+ \lambda(u)\frac{\del  u}{\del  x}
= \eta \frac{\del ^2 u}{\del x^2},
\eeq
where $\lambda(u)$ is a sufficiently smooth function of $u$, and $\eta$ is
a positive constant.
Following \cite{goldenfeld2}, we consider a small amplitude wave in the
background of constant solution $u_0$;
\beq
u= u_0+\eps u_1 + \eps^2 u_2 + \cdots.
\eeq
Here  $\eps$ denotes the amplitude of the wave.  $\lambda(u)$ is
  expanded as
\beq
\lambda(u)= \lambda_0 + \eps \lambda '(u_0)u_1+ \cdots,
\eeq
 where
$\lambda_0\equiv \lambda(u_0)$.
We further assume that the dissipation is weak in the sense $\eta \sim \eps$;
 we write as $\eta = \mu \eps$, where $\mu =O(1)$.
Then equating the coefficients of $\eps ^n$ ($n=0, 1, 2 ,...$), we have
\beq
\d _tu_1 + \lambda_0\d _xu_1 &=&0, \nonumber \\
\d _tu_2 +\lambda_0\d _xu_2 &=& -\lambda'_0u_1\d _xu_1 +\mu \d _x^2u_1,
\eeq
and so on. Here, $\lambda'_0=\lambda'(u_0)$.

It is now convenient to introduce the new variable $\xi\equiv x-\lambda_0 t$,
the characteristic direction of the unperturbed equation;  we describe the
 solutions in terms of $(\xi, t)$.
Then
\beq
\d _tu_1&=&0, \nonumber \\
\d _tu_2 &=& -\lambda'_0u_1\d _{\xi}u_1 +\mu \d _{\xi}^2u_1,
\eeq
and so on.
One readily gets
\beq
u_1(\xi, t)&=&F_0(\xi), \nonumber \\
u_2(\xi, t)&=& (t-t_0)( -\lambda'_0F_0(\xi)\d _{\xi}F_0(\xi) +
\mu \d _{\xi}^2F_0(\xi)),
\eeq
where $F_0(\xi)$ is an arbitrary function of $\xi$.
 We note that $u_2$ is a secular term.\footnote{
 One might have added another arbitrary function
$G_0(\xi)$of  $\xi$ to $u_2$. However, the effect of $G_0$ could be
 renormalized away to $F_0$.}

We have thus an approximate solution to Eq. (2.1)
\beq
u(\xi, t; t_0)=u_0 + \eps F_0(\xi) +
\eps ^2 (t-t_0)( -\lambda'_0F_0(\xi)\d _{\xi}F_0(\xi) +
\mu \d _{\xi}^2F_0(\xi)) + O(\eps ^3),
\eeq
where we have made it explicit
that $u$ is dependent on an arbitrary time $t_0$. We stress that
this (approximate) solution is only valid for $t$ around $t_0$.

Now, geometrically speaking, we have a family of surfaces $S_{t_0}$
represented
 by $u(\xi, t;t_0)$ with a parameter $t_0$.
Let us obtain the envelope surface $E$  of this family of surfaces,
 following  the classical theory of envelopes given in Appendix A:
 we represent $E$ by $u_{_E}(\xi, t)$.

We first note that  $F_0(\xi)$ may be  functionally dependent
 on the arbitrary initial time  $t_0$:
\beq
u(\xi, t; t_0)=u_0 + \eps F_0(\xi, t_0) +
\eps ^2 (t-t_0)( -\lambda'_0F_0(\xi, t_0)\d _{\xi}F_0(\xi, t_0) +
\mu \d _{\xi}^2F_0(\xi, t_0)) + O(\eps ^3).
\eeq
It is natural to set the  tangent curve $C_{t_0}$ of $E$ and $S_{t_0}$
to lie along  $t=t_0$-line, which is parallel to the characteristic
direction ($\xi$-direction):
\beq
{\rm C}_{t_0}:\ \ \ t=t_0,\ \ \ u=u(\xi, t_0, t_0),\ \ \ \xi=\xi.
\eeq
Then the envelope equation reads
\beq
\frac{\d u}{\d t_0}=0,\ \ \ {\rm and}\ \ t_0=t,
\eeq
which leads to
\beq
\del _tF(\xi, t) + \eps \lambda'(u_0)F\del _{\xi}F = \eta \del ^2_{\xi}F.
\eeq
This is Burgers equation which is usually obtained in the reductive
 perturbation theory.
Then
\beq
u_{_E}(\xi, t)= u(\xi, t; t)= u_0 + \eps F(\xi, t).
\eeq

Now one may wonder if  $u_{_E}(\xi, t)$ satisfies Eq. (2.1),
 although $u(\xi, t; t_0)$
 does up to $O(\eps ^3)$ for any $t_0$. We remark that the question is not
trivial because $u_{_E}(\xi, t)\equiv u(\xi, t; t)$. We shall show that the
 answer is
 yes.  In fact, the time derivative of $u_{_E}(\xi, t)$ at $t= \forall t_0$
 coincides with $\del u(\xi, t; t_0)/\del t$ at $t=t_0$;
\beq
\frac{\del u_{_E}(\xi, t)}{\del t}\Biggl\vert_{t=t_0}&=&
\frac{\del u(\xi, t; t_0)}{\del t}\Biggl\vert_{t=t_0}+
\frac{\del u(\xi, t_0; t'_0)}{\del t'_0}\Biggl\vert_{t'_0=t_0},\nonumber \\
 \ \ \ &=& \frac{\del u(\xi, t; t_0)}{\del t}\Biggl\vert_{t=t_0},
\eeq
where Eq. (2.10) has been used. Furthermore, needless to say,
 $u_{_E}(\xi, t_0)= u(\xi, t_0, t_0)$. Thus, at $t=\forall t_0$,
\beq
\frac{\del u_{_E}}{\del t}+
\lambda(u_E)\frac{\del u_{_E}}{\del x}= \eta \frac{\d ^2 u_{_E}}{\d x^2}
 + O(\eps^3).
\eeq
Thus one sees that our envelope function $u_{_E}$ satisfies the Eq.(2.1)
uniformly for  $t$ in the global range.

 A comment is in order here:  one may say that to get the approximate but
global solution
 $u_{_E}$ from the local solution Eq.(2.8), we have utilized the fact that
 the uniqness of the solution of the Cauchy problem is violated when
 the initial data are given along the characteristic manifold (curve);
 see Appendix A for characteristic manifolds(curves).

\setcounter{equation}{0}
\section{One-dimensional Swift-Hohenberg equation}
\renewcommand{\theequation}{\thesection.\arabic{equation}}

In this section, we deal with the one-dimensional
Swift-Hohenberg equation\cite{swift,cross}:
\beq
\hat{L}_1u=\eps u -u^3, \ \ \ \ \hat{L}_1\equiv \del _t+(\del_x^2 +k^2)^2,
\eeq
 where $\eps$ is a small parameter.
We shall show that the envelope of  local solutions of
 Eq. (3.1)  satisfies
 the time-dependent Ginzburg-Landau equation.

Following \cite{goldenfeld1}, we scale $u$ as
\beq
u=\sqrt{\eps}\phi.
\eeq
Then $\phi$ satisfies
\beq
\hat{L}_1\phi=\eps (\phi -\phi^3).
\eeq
We solve this equation in the perturbation theory, expanding
$\phi$ as $\phi = \phi_0 + \eps \phi_1 + ...$; we have
\beq
\hat{L}_1\phi_0=0, \ \ \ \hat{L}_1\phi_1=\phi_0-\phi_0^3 ,
\eeq
and so on.
As the 0-th order solution, we take
\beq
\phi_0(x,t)= Ae^{ikx} + {\rm c.c.},
\eeq
where c.c. denotes the complex conjugate.
Then the 1st-order equation reads
\beq
\hat{L}_1\phi_1= {\cal A}e^{ikx} - A^3e^{3ikx} + {\rm c.c.},
\eeq
where
\beq
 {\cal A}\equiv A(1-3\vert A\vert ^2).
\eeq
A special solution to Eq. (3.6) is found to be\cite{goldenfeld2}
\beq
\phi_1 = {\cal A}\{\mu_1(t-t_0) - \frac{\mu_2}{8k^2}(x^2-x_0^2)\}
 e^{ikx} + \frac{A^3}{64k^3}e^{3ikx} + {\rm c.c.},
\eeq
 with $\mu_1 +\mu_2=1$. We note that the secular terms have
 appeared in $\phi_1$ because of the first term in r.h.s. of Eq. (3.6).

Now one may say that we have a family of surfaces $S_{t_0x_0}$
represented by
\beq
\phi(x, t; x_0, t_0)&=&[Ae^{ikx} +
\eps {\cal A}\{\mu_1(t-t_0) - \frac{\mu_2}{8k^2}(x^2-x_0^2)\}
 e^{ikx} + \eps\frac{A^3}{64k^3}e^{3ikx}]\nonumber \\
 \ \ \ & \ \ \ & + {\rm c.c.},
\eeq
parametrized by $t_0$ and $x_0$.  Let us obtain the envelope of the family
of the surfaces in  two steps, by assuming that the amplitude $A$ is
 dependent on $x_0$ and $t_0$. First, fixing $x_0$, we obtain the envelope
 $E_{1}$ of the surfaces with $t_0$ being the parameter. The resultant
envelope $E_{x_0}$ has the parameter $x_0$; then we obtain the envelope of
the  family of the surfaces $E_{1}$.

The first step is achieved by setting
\beq
\frac{\del \phi}{\del t_0}=0, \ \ \  {\rm with }\ \  \ t_0=t.
\eeq
This is the condition to get the envelope that has the common tangent curve
with  $S_{x_0t_0}$ along $x$-direction.  Thus we have the equation for
$A(x_0, t)$;
\beq
\del_t A=\mu_1\eps {\cal A} + O(\eps^2).
\eeq
Then the envelope $E_{1}$ is represented by
\beq
\phi_{E_{1}}(x, t;x_0)&\equiv& \phi(x, t;x_0, t_0=t),\nonumber \\
 \ \ \ &=&  [A(x_0, t)e^{ikx} -
\eps {\cal A}(x_0,t)\frac{\mu_2}{8k^2}(x^2-x_0^2)
 e^{ikx} + \eps\frac{A(x_0,t)^3}{64k^3}e^{3ikx}]\nonumber \\
 \ \ \ & \ & + {\rm c.c.}.
\eeq

The envelope $E$ of the family of the surfaces given by Eq. (3.12)
  is obtained as follows;
\beq
\frac{\del \phi_{E_{1}}}{\del x_0}=0 , \ \ \ \
 {\rm with}\ \ \  x_0=x,
\eeq
which leads to
\beq
\del_xA(x,t)= - \mu_2\eps {\cal A}\frac{x}{4k^2} +
              O(\eps^2).
\eeq
Here we have utilized the fact that $\del_xA$ is $O(\eps)$.
Differentiating this equation with respect to $x$, we get
\beq
\del_x^2A(x,t)= -\eps \frac{\mu_2}{4k^2} {\cal A} + O(\eps^2).
\eeq
Here we note again that $\del_xA$ is $O(\eps)$.

Now combining Eq.'s (3.10) and (3.14), we have
\beq
\del_tA = 4k^2 \del_x^2A + \eps A(1- 3\vert A\vert^2),
\eeq
 up to $O(\eps^2)$.
The envelope is represented by
\beq
\phi_E(x,t)&\equiv& \phi_{E_{1}}(x,t;x_0=x),\nonumber \\
 \ \ \ &=& \Big[A(x, t)e^{ikx} +\eps \frac{A(x, t)^3}{64k^4}e^{3ikx}
              \Big] + {\rm c.c.}.
\eeq

A comment is in order here: To derive Eq.(3.15), we started from
 the first-order differential equation Eq.(3.13) with respect to $x_0$.
 One could start from the second-order differential equation as is
 done in \cite{goldenfeld2};
\beq
\frac{\del^2\phi}{\del x_0^2}\Biggl\vert _{x_0=x}=0,
\eeq
which gives Eq.(3.15) as an exact relation without the remainder of
 $O(\eps^2)$.  We remark that the geometrical meaning of
 Eq.(3.18) is not clear in contrast to Eq.(3.13).  In our formulation,
 Eq.(3.18) is derived as an approximate relation, as is shown below.

Now one may ask the question as to  the relation of $\phi_E(x, t)$ given
 in Eq. (3.17) and the original equation Eq.(3.1): The answer is that
 $\phi_E(x,t)$ satisfies Eq.(3.1) up to $O(\eps^2)$ uniformly
 in the global domain due to the very envelope conditions Eq.(3.10)
 and Eq.(3.13).  In fact, for $\forall t=t_0$,
\beq
\frac{\del \phi_E}{\del t}\Biggl\vert _{t=t_0}=
\frac{\del \phi(t,t_0)}{\del t}\Biggl\vert _{t=t_0}+
\frac{\del \phi (t, t_0)}{\del t_0}\Biggl\vert _{t=t_0}=
\frac{\del \phi(t,t_0)}{\del t}\Biggl\vert _{t=t_0},
\eeq
on account of Eq.(3.10).   Similary, for $\forall x= x_0$,
one can easily verify that
\beq
\frac{\del^2\phi_E}{\del x^2}\Biggl\vert _{x=x_0}=
\frac{\del^2\phi}{\del x^2}\Biggl\vert _{x=x_0} + O(\eps^2),
\eeq
 on account of Eq.'s (3.12), (3.13) and (3.18).  For instance,
 $\del^2\phi_E/\del x\del x_0\vert _{x=x_0}=O(\eps^2).$

\setcounter{equation}{0}
\section{Damped Kuramoto-Shivashinsky equation}
\renewcommand{\theequation}{\thesection.\arabic{equation}}

In this section, we deal with the (one-dimensional)
 damped Kuramoto-Shivashinsky equation, which is given by
\beq
\hat{L}_1 u= \eps u  -u\del_xu,
\eeq
where $\hat{L}_1$ is defined in Eq.(3.1). This equation is not examined
 by the Illinois group.

By scaling $u$ as
\beq
u=\sqrt{\eps}\phi,
\eeq
we have\footnote{If one scales $u$ as $u=\eps \phi$, one will get
 a different equation from that given below.}
\beq
\hat{L}_1{\phi} = - \eps^{1/2}\phi\del_x\phi + \eps \phi.
\eeq

We first try to solve this equation by the perturbation theory
with the expansion
\beq
\phi = \phi_0 + \eps^{1/2}\phi_1 + \eps \phi_1 + ... .
\eeq
The equations for $\phi_0, \phi_1, \phi_2 ... $ are found to be
\beq
\hat{L}_1\phi_0=0, \ \ \ \hat{L}_1\phi_1 =-\phi_0\del_x\phi_0, \ \ \
 \hat{L}_1\phi_2 = \phi_0 -(\phi_0\del_x\phi_1+\phi_1\del_x\phi_0),
\eeq
 and so on.

As the 0-th order solution, we take
\beq
 \phi_0= Ae^{ikx} + {\rm c.c},
\eeq
where $A$ is a constant. Then $\phi_1$ is found to be
\beq
\phi_1= -\frac{i}{9k^2}A^2e^{2ikx} + {\rm c.c.}.
\eeq
Thus the equation for $\phi_2$ reads
\beq
\hat{L}_1\phi_2= A(1 -\frac{\vert A\vert^2}{9k^2})e^{ikx}-
                \frac{A^3}{3k^3}e^{3ikx} + {\rm c.c.},
\eeq
 which is in a similar form with Eq.(3.6). One easily gets for a
 special solution to this equation
\beq
\phi_2= A(1 -\frac{\vert A\vert^2}{9k^2})\{
       \mu_1(t-t_0)- \frac{\mu_2}{8k^2}(x^2 - x_0^2)\}e^{ikx}
            -\frac{A^3}{192k^6}e^{3ikx} + {\rm c.c.}.
\eeq
Thus we reach the solution in the perturbation theory up to
$O(\eps^{3/2})$,
\beq
\phi(x,t;x_0,t_0)&=&\Big[A\e^{ikx} -i\eps^{1/2}\frac{A^2}{9k^3}e^{2ikx}
  +\eps A(1 -\frac{\vert A\vert^2}{9k^2})\{
       \mu_1(t-t_0)- \frac{\mu_2}{8k^2}(x^2 - x_0^2)\}e^{ikx}\nonumber \\
    \ \ \ & \ \ \ & \ \ -\eps\frac{A^3}{192k^6}e^{3ikx}\Big]
          +{\rm c.c.}.
\eeq

Now we have a family of surfaces $S_{x_0t_0}$ in $x$-$t$-$\phi$ plane
represented by $\phi(x,t;x_0,t_0)$ with $x_0$ and $t_0$ being the parameters.
We repeat the procedure of the previous section to obtain the envelope $E$
 of the family of the surfaces: We first note that $A$ may depend on
$x_0$ and $t_0$, i.e., $A=A(x_0, t_0)$. Then fixing $x_0$,
 we first obtain the envelope of
 $S_{x_0t_0}$ with $t_0$ being the parameters.  The envelope $E_1$ is
 obtained by
\beq
\frac{\del \phi}{\del t_0}=0 , \ \ \ {\rm with} \ \ \ t_0=t,
\eeq
where we have assumed that the tangent curve (the characteristic curve)
 is along $x$-direction by setting $t_0=t$.  The above equation  leads to
\beq
\del _tA(x_0, t)=\eps\mu_1A(1 -\frac{\vert A\vert^2}{9k^2}) +
                  O(\eps^{3/2}).
\eeq
With $A$ satisfying this equation, the envelope $E_1$ is represented by
\beq
\phi_{E_1}(x, t; x_0)&\equiv &\phi(x_0,t_0; x_0, t_0=t)\nonumber \\
   \ \ \ \ &=& \Big[A\e^{ikx} -i\eps^{1/2}\frac{A^2}{9k^3}e^{2ikx}
  -\eps A(1 -\frac{\vert A\vert^2}{9k^2})\frac{\mu_2}{8k^2}
(x^2 - x_0^2)e^{ikx}\nonumber \\
    \ \ \ & \ \ \ & \ \ -\eps\frac{A^3}{192k^6}e^{3ikx}\Big]
          +{\rm c.c.}.
\eeq
This function can be regarded as representing a family of surfaces with
 $x_0$  being the parameter.  The envelope $E$ of this family of surfaces
 are obtained by setting
\beq
\frac{\del \phi_{E_1}}{\del x_0}=0, \ \ \ \ {\rm with }\ \ \ x_0=x,
\eeq
which leads to
\beq
\del _xA(x,t)=
          -\eps \frac{\mu_2}{4k^2}A(1 -\frac{\vert A\vert^2}{9k^2})x
                      + O(\eps ^{3/2}).
\eeq
Here we have utilized the fact that $\del _xA\sim O(\eps)$.  Further
 differentiating with respect to $x$, one has
\beq
\del_x^2 A=  -\eps \frac{\mu_2}{4k^2}A(1 -\frac{\vert A\vert^2}{9k^2})
                      + O(\eps ^{3/2}).
\eeq
Combining Eq.'s (4.12) and (4.16), one sees that the amplitude satisfies
equation
\beq
(\del_t - 4k^2\del _x^2)A=\eps A(1 -\frac{\vert A\vert^2}{9k^2}),
\eeq
 up to $O(\eps^2)$.

With this amplitude, the envelope $E$ is given by
\beq
\phi_{E}(x,t)&\equiv &\phi(x,t;x_0=x),\nonumber \\
 \ \ \ \ &=& Ae^{ikx}-i\eps^{1/2}\frac{A^2}{9k^2}e^{2ikx}
              - \eps \frac{A^3}{192k^6}e^{3ikx} + {\rm c.c.}.
\eeq

A couple of comments are in order here:

(1)\ Eq.(4.16) could be
 obtained  as an exact relation by imposing the condition that
\beq
\frac{\del ^2 \phi}{\del x_0^2}\Biggl\vert _{x_0=x}=0.
\eeq
 We must note, however, that the geometrical meaning of this condition is
not clear.  In contrast, in our formulation, the second derivative is
 evaluated to be
\beq
\frac{\del ^2 \phi}{\del x_0^2}\Biggl\vert _{x_0=x}= O(\eps^2).
\eeq

(2) As was done in the preceding section, one can easily show that
  $\phi_E(x,t)$ satisfies the original equation Eq.(4.1) up to
$O(\eps^2)$ but uniformly in the global domain.

\setcounter{equation}{0}
\section{Two-dimensional Swift-Hohenberg equation}
\renewcommand{\theequation}{\thesection.\arabic{equation}}

In this section, we deal with the two-dimensional Swift-Hohenberg
 equation given by
\beq
\hat{L}_2u=\eps u - u^3, \ \ \ \
       \hat{L}_2= \del_t + (\del_x^2 + \del_y^2+ k^2)^2.
\eeq
Scaling $u$ as
\beq
u=\sqrt{\eps}\phi,
\eeq
we have
\beq
\hat{L}_2\phi=\eps (\phi -\phi^3).
\eeq
We first solve this equation in the naive perturbation theory expanding
$\phi = \phi_0 + \eps \phi_1 + \eps^2\phi_2 + ... $;
\beq
\hat{L}_2\phi_0=0,\ \ \ \
   \hat{L}_2 \phi_n = \phi_{n-1}- \phi_{n-1}^3, \ \ \ (n=1, 2, ...).
\eeq
If we assume the roll solution along $y$-axis for the zero-th order
  equation,
\beq
\phi_0= Ae^{ikx} + {\rm c.c.},
\eeq
then the solution up to $O(\eps^2)$ is found to be \cite{goldenfeld3}
\beq
\phi(x,t;x_0,t_0)&=& \Big[ A+
        \eps \{\mu_1(t-t_0) -\mu_2\frac{x^2-x_0^2}{8k^2} +
               \mu_3\frac{xy^2 - x_0y_0^2}{8ik}+
                \mu_4\frac{y^4 -y_0^4}{4!}\}{\cal A}
                   \Big]e^{ikx} \nonumber \\
 \ \ \ \ & \ \ & \ \ + {\rm c.c.},\ \ \ {\rm with } \ \ \
            {\cal A}\equiv A(1-3\vert A\vert^2),
\eeq
where $\sum_{i=1\sim 4}\mu_i=1$ and
$x_0, y_0$ and $t_0$ are arbitrary constants. Here we have omitted the
 terms that do not give rise to secular terms.

Now one may regard that we have a family of ``surfaces'' in the
$x$-$y$-$t$-$\phi$ space with $x_0, y_0$ and $t_0$ being the parameters.
Let us obtain the envelope of this ``surfaces'' in the three steps by
 noting that $A$ may be dependent on $x_0, y_0$ and $t_0$:
 First we regard that only $t_0$ is the parameter of the ``surfaces''
 with both $x_0$ and $y_0$ fixed.  The condition reads
\beq
\frac{\del \phi}{\del t_0}=0, \ \ \ \ {\rm t_0=t}.
\eeq
This condition may be also regarded as the one for the envelope curve
 of a family of curves in $t$-$\phi$ plane with $x$ and $y$ being fixed.
The condition leads to
\beq
\del _t A(x_0, y_0, t)= \eps\mu_1{\cal A} + O(\eps^2).
\eeq
Inserting this solution, we have the envelope
\beq
\phi_{E_1}(x, y,t; x_0, y_0)&\equiv &\phi(x, y, t; x_0, y_0, t_0=t)
          \nonumber \\
 \ \ \ &=& \Big[ A(x_0, y_0, t)+
        \eps \{-\mu_2\frac{x^2-x_0^2}{8k^2} +
               \mu_3\frac{xy^2 - x_0y_0^2}{8ik}+
                \mu_4\frac{y^4 -y_0^4}{4!}\}{\cal A}
                   \Big]e^{ikx}\nonumber \\
 \ \ \ \ & \ & \ \ \  + {\rm c.c.},
\eeq
 which we may regard as a family of curves in $x$-$\phi$ plane with
$x_0$ being the parameter where $y$ and $y_0$ are fixed.  The envelope
$E_2$ for this family of curves is given by setting
\beq
\frac{\del \phi_{E_1}}{\del x_0}=0, \ \ \ \ {\rm  with } \ \ \ x_0=x,
\eeq
which leads to
\beq
\del _x A(x, y_0,t)= -\eps
                  (\frac{\mu_2}{4k^2}x -\frac{\mu_3}{8ik}y_0^2)
                  {\cal A} + (\eps^2).
\eeq
Accordingly, the envelope is represented by
\beq
\phi_{E_2}(x, y,t; y_0)&\equiv &\phi_{E_1}(x, y, t; x_0=x, y_0)\nonumber \\
 \ \ \ &=& \Big[ A(x, y_0, t)e^{ikx}+
        \eps \{ \mu_3x\frac{y^2 - y_0^2}{8ik}+
                \mu_4\frac{y^4 -y_0^4}{4!}\}{\cal A}e^{ikx}
                   \Big] + {\rm c.c.},
\eeq
which is regarded as representing a family of curves in $y$-$\phi$ plane.
The envelope of $\phi_{E_2}$ is obtained as usual by setting
\beq
\frac{\del \phi_{E_2}}{\del y_0}=0, \ \ \ \ {\rm with} \ \ \ y_0=y,
\eeq
which leads to
\beq
\del _y A(x, y, t)= \eps(\frac{\mu_3}{4ik}xy + \frac{\mu_4}{3!}y^3)
                     {\cal A} + O(\eps^2).
\eeq

Differentiating Eq.'s (5.11) and (5.14), we have
\beq
\del_x^2A(x,y,t)&=& -\eps  \frac{\mu_2}{4k^2}{\cal A} +
                   O(\eps^2),\\
\del_x\del^2_{y}A(x,y,t)&=& -\eps  \frac{\mu_3}{4ik}{\cal A} +
                   O(\eps^2), \\
\del^4_yA(x,y,t)  &=& \eps  \mu_4{\cal A} +
                   O(\eps^2),
\eeq
where we have utilized the fact that
\beq
\del_xA\sim O(\eps), \ \ \ \del_yA\sim O(\eps).
\eeq
Combining these equation together with Eq.(5.8), we finally reach
\beq
\del_tA -4k^2\del_x^2A + 4ik \del_x\del_y^2A + \del^4_yA=
        \eps A(1 - 3\vert A\vert ^2),
\eeq
 up to $O(\eps^2)$.

With this amplitude, the envelope is given by
\beq
\phi_E(x, y, t)&=&\phi_{E_1}(x, y, t; y_0=y),\nonumber \\
 \ \ \ &=& A(x, y, t) e^{ikx} + {\rm c.c.}.
\eeq

A few comments are in order here:

(1) By imposing that the any order of the differentiation
    of $\phi$ with respect to $x_0$ and $y_0$ should vanish, one
    could get  Eq.'s (5.15-17)  as exact relations
      \cite{goldenfeld3}, although the geometrical meaning of
     these conditions are unclear.  In our formulation, instead,
    the following  are derived as  approximate relations
\beq
\frac{\del^2\phi_{E1}}{\del^2x_0}\Biggl\vert_{x_0=x}=O(\eps^2),\ \ \
\frac{\del^3\phi_{E2}}{\del x\del^2y_0}\Biggl\vert_{y_0=y}=O(\eps^2),\ \ \
\frac{\del^4\phi_{E2}}{\del^4y_0}\Biggl\vert_{y_0=y}=O(\eps^2).
\eeq

(2) It can be shown that $\phi_E$ satisfies Eq.(5.1) up to $O(\eps^2)$
     but uniformly in the global domain owing to the above relations together
     with Eq. (5.8).

\section{A brief summary and concluding remarks}

We have shown in the present paper that the RG method for global
analysis can be formulated for partial differential equations, too.
  An interesting equation which is treated by the Illinois
 group\cite{goldenfeld1} but left
 untouched in the text is Barenblatt's equation.
 A complete global solution to this equation is
 not given in the RG method\cite{goldenfeld1} but only the anomalous
 exponent for the long-time behavior is obtained. For completeness, we
  show that the same result for the anomalous exponent can be obatined
 in our envelope theory in Appendix B.
 Thus together with our previous
 paper\cite{kunihiro} where ordinary differential equations are discussed,
 we have shown that almost all the classes of problems treated in the
RG method  by the Illinois group\cite{goldenfeld1,goldenfeld2,goldenfeld3}
 are nicely formulated on the basis of the classical theory of envelopes
 without recourse to
 the notion of the renormalization group. Actually, this is
 natural because the resulting equations indeed describe the
amplitudes of the nonlinear waves as given by
  the solutions of the nonlinear
 equations.

It is already indicated\cite{goldenfeld1,goldenfeld2}  that
the renormalizability  is equivalent to the solvability of
equations\cite{kuramoto}. It would be interesting to apply
 the theory developed here to systems of equations\footnote{The RG method or
 the envelope theory developed in \cite{kunihiro} and here is
 of course applicable to systems of ODE's\cite{kunihiro3}.} and see
 possible geometrical meaning of the solvability condition.

In deriving the whole envelopes of local solutions, we have taken a multi-step
 approach in sections 3, 4 and 5.  It is interesting that a kind of
 multi-step approach is also proposed for improving the effective potentials
 in quantum field theories with multi-scales; see the paper by
 Ford\cite{kugo}. It may imply that our approach that identifies the
 RG equation with the envelope equation naturally leads to
  the effective potentials as given by Ford when applied to quantum
 field theories.\cite{kunihiro4}

\vspace{2cm}
{\cl {\large{\bf Acknowledgements}}}

The author acknowledges G. C. Paquette, who gave
 lectures on the RG method at  Ryukoku University in June 1994.
He is also grateful to R. Kobayashi, who organized the seminar where
 these lectures were given.
He thanks Y. Morita, the conversation with whom motivated
 the author to think about the mathematical structure of the RG method
 seriously.
The author is grateful to M. Yamaguti for his interest in this work,
 encouragement and  useful comments on Cauchy's problem and Burgers
equation.
He also acknowledges Y. Oono for his correspondence and
 sending the preprint\cite{goldenfeld3} prior to publication.
 The author thanks J. Matsukidaira for discussions on applications
 of the envelope theory to systems of ODE's.
 A part of the present work was completed while the author stayed at
 Brookhaven Natinal Laboratory(BNL) as a summer-program visitor in August
 3-14, 1995. He thanks BNL and especially R. Pisarski for their hospitality.

\newpage

\setcounter{equation}{0}
{\large {\bf Appendix A}}
\renewcommand{\theequation}{A.\arabic{equation}}

In this appendix, we  give a short review of the classical theory of
envelopes\cite{courant}.
We shall first consider   envelopes in three-dimensional space. i.e.,
envelope  surfaces. Then the extension to higher dimensional cases
 will be briefly described.\footnote{In \cite{kunihiro}, a review is
 given on how to construct  envelope curves.}

Let $\{S_{\tau}\}_{\tau}$ be a family of surfaces  parametrized by
$\tau$ in the $x$-$y$-$u$ space; here $S_{\tau}$ is  represented by the
 equation
\beq
\Phi(x, y, u; \tau)=0.
\eeq
 We suppose
 that $\{S_\tau\}_{\tau}$ has the envelope $E$, which is represented by
the equation
\beq
\Psi(x, y, u)=0.
\eeq
The problem is to obtain $\Psi(x, y, u)$ from $\Phi(x, y, u; \tau)$.

Let $E$ and  $S_{\tau}$ has a common tangent plane on  a common curve
$C_{\tau}$;
 with $\tau$ being varied, $\{C_{\tau}\}_{\tau}$ forms $E$. $C_{\tau}$
is called a {\em characteristic curve}.
The necessary condition for $S_{\tau}$ to have an envelope $E$ is given
as follows.
Let $C_{\tau}$ be represented by
 $x=x(\sigma, \tau),\ y=y(\sigma, \tau), \ u=u(\sigma, \tau)$, with $\sigma $
 being a  parameter.  We assume that
\beq
{\rm rank}\pmatrix{x_{\sigma}\ y_{\sigma}\ u_{\sigma} \cr
                  x_{\tau}\ y_{\tau}\ u_{\tau} \cr }=2,
\eeq
 in order for $\{C_{\tau}\}_{\tau}$ to give a non-degenerate surface $E$.
Now, since $C_{\tau}$ is on $S_{\tau}$,\\
 $\Phi(x(\sigma, \tau), y(\sigma, \tau), u(\sigma, \tau); \tau)=0.$
Differentiating this equation with respect to $\tau$, one has
\beq
\Phi_xx_{\tau} + \Phi_yy_{\tau}+\Phi_uu_{\tau}+ \Phi_{\tau}=0,
\eeq
where $ \Phi_x\equiv \del \Phi/\del x$ and so on.
 On the other hand, since
$(x_{\tau}, y_{\tau}, u_{\tau})\equiv {\bf t}_{\tau}$ is a tangent vector
both  of $E$ and $S_{\tau}$, and
$(\Phi_x, \Phi_y, \Phi_u)\equiv {\bf n}_{\tau}$ is a
normal vector of $S_{\tau}$,
\beq
{\bf n}_{\tau}\cdot {\bf t}_{\tau}=
\Phi_xx_{\tau} + \Phi_yy_{\tau}+\Phi_uu_{\tau}=0.
\eeq
Combining Eq.'s (A.4) and (A.5), one has
\beq
\Phi_{\tau}\equiv \frac{\del \Phi(x, y, u;\tau)}{\del \tau}=0,
\eeq
as a necessary condition for $\{S_{\tau}\}$ to have an envelope $E$.
Thus, solving Eq. (A.6) for $\tau=\tau(x, y, u)$, one finally gets
\beq
\Psi(x, y, u)= \Phi(x, y, u; \tau(x, y, u))=0.
\eeq

Accordingly, the  characteristic curve is given by the conditions;
\beq
\Psi(x, y, u)=0 \ \ \ {\rm and}\ \ \
\tau(x, y, u)= {\rm const}.
\eeq
Note that the second equation gives a constraint on $x, y$ and $u$.

We remark that the sufficient conditions for the existence of an envelope
 is supplemented by \cite{courant}
\beq
\Phi_{\tau \tau}=0.
\eeq

When $S_{\tau}$ is given by $u=\varphi(x, y; \tau)$,
the condition Eq. (A.6) is reduced to
\beq
\frac{\del \varphi(x, y; \tau)}{\del \tau}=0,
\eeq
 which gives  $\tau$ as a function of $x$ and $y$. Thus we get for
 the  envelope
\beq
u= \varphi_{_E}(x,  y)=\varphi(x, y; \tau(x, y)).
\eeq
The  characteristic  curve is accordingly given  by
\beq
u= \varphi_{_E}(x,  y)\Bigl\vert_{\tau(x, y)={\rm const}.}.
\eeq

As an example, let
\beq
u=\varphi(x, y, \tau)\equiv {\rm exp}(\tau y)( 1 + y(x- \tau)) +
       {\rm exp}(-x).
\eeq
 Then the equation $\del \varphi/\del \tau=0$ gives
\beq
\tau = x,
\eeq
except at $y\not=0$.
Thus the envelope is given by
\beq
u={\rm exp}(x y) + {\rm exp}(-x).
\eeq
 The  characteristic curve  C$_{\tau}$ is  given by
\beq
x=\tau= {\rm constant} \equiv x_0, \ \ \ {\rm and}\ \ \
u={\rm exp}(x_0\ y) + {\rm exp}(-x_0).
\eeq
 We remark that
$\del ^2 \varphi/\del \tau^2\not=0$ provided that $y\not=0$.

The theory  can be extended to the envelope of
a family of hyper-surfaces $\{S_{\tau}\}_{\tau}$
in the $(n+1)$-dimensional space ${\bf R}^{n+1}$.
 Let $\{S_{\tau}\}_{\tau}$  be represented by the equation
$\Phi(x_1, x_2, \cdots , x_n, u; \tau)=0$  with a parameter set
$\tau =(\tau_1, \tau_2, \cdots , \tau_{n-1})$.  Then
the function representing the envelope is given by eliminating parameters
 $\tau_i$  $(i= 1, 2, \cdots , n-1)$ from
\beq
\Phi(x_1, x_2, \cdots , x_n, u; \tau)=0, \ \ \frac{\del \Phi}{\del \tau_i}=0,
 \ \ \ (i= 1, 2, \cdots , n-1).
\eeq

A few comments are in order here:

(i)\ \   The envelope
of a family of surfaces   has usually an improved  global nature
compared with  the surfaces  in the family.
 So it is natural that the theory of envelopes may have  some power
 for global analysis.

(ii)\ \
It should be stressed here that Eq.(A.6), which we call the envelope equation,
 has the same form as renormalization group (RG) equations.  Rather,
it should be said conversely;
 RG equations in general are  envelope equations\cite{kunihiro}.
This is the reason why RG equations ``improve'' things; one should note
 that the improvement by an RG equation usually means that a function
 with a better global nature is constructed from functions with a local
 nature only valid around the renormalization point.

(iii)\ \ Let $u=\varphi(x, y; \sigma, \tau)$ be a complete solution
 of a PDE
\beq
F(x, y, u, p, q)=0, \ \ \ \ (p\equiv \del u/\del x,\ \
q\equiv \del u/\del y),
\eeq
where  $\sigma $ and $\tau$ are constant.
Assuming  a functional dependence of $\sigma$ on $\tau$, i.e.,
 $\sigma = A(\tau)$,
 one can obtain the envelope as follows;
\beq
\frac{d u}{d \tau}=\frac{dA}{d\tau}\frac{\del u}{\del \sigma}
 + \frac{\del u}{\del \tau}=0,
\eeq
from which $\tau$ is given as a function of $x$ and $y$, namely,
$\tau=\tau(x, y)$.  Then one sees that the envelope function
$u=\varphi_{_E}(x, y; [A])=\varphi(x, y, A(\tau(x, y)), \tau(x, y))$ is
 the general solution to $ F(x, y, u, p, q)=0$.

\setcounter{equation}{0}
\begin{large}
\cl{\bf Appendix B}
\end{large}
\renewcommand{\theequation}{B.\arabic{equation}}

In this Appendix, we examine a non-linear partial differential equation
 which has an anomalous exponent in the long-time behavior of its solutions;
\beq
(\partial _t - \half \partial_x^2)u(t,x)=\theta(-\partial_tu)\partial_x^2u.
\eeq
This is called Barlenblatt's equation.\cite{baren,goldenfeld1}
This equation  is a nonlinear PDE with the step
 function in r.h.s.  Without this term ($\eps =0$),
 the equation is a simple diffusion
 equation and has a self-similar form after a long time;
$u(x, t)\sim t^{-1/2}f(x^2/t)$.  However, with $\eps\not=0$, the
long-time behavior is found to be
\beq
u(x, t)\sim t^{-(\frac{1}{2}+ \alpha)}F(\frac{x^2}{t}).
\eeq
The anomalous exponent $\alpha$ can be interpreted as an anomalous dimension
\cite{goldenfeld1}. The appearance of the anomalous
 dimension is due to the fact that the scale $l$ with a spatial dimension
which
 characterizes the initial distribution of $u(x, 0)$ can not be neglected
 even in the long-time limit $t\rightarrow \infty$ in contrast to the
  case of the simple diffusion equation \cite{baren}.
 We are interested in determining the anomalous dimension $\alpha$ in the
 perturbation theory.

To solve the problem, we shall  follow
  \cite{goldenfeld1} for a while.
 First we  convert the equation to an integral equation;
\beq
u(x, t)& =& \int _{-\infty}^{\infty}dy G(x-y, t)u(y,0)\nonumber \\
 \ \ \ &\ & +\frac{\eps}{2}\int_0^tds\int _{-\infty}^{\infty}dy
  G(x-y, t-s)\theta[-\del_s u(y,s)]\del^2_yu,
\eeq
where
\beq
G(x, t)=\frac{1}{\sqrt{2\pi t}}{\rm e}^{-\frac{x^2}{2t}}
\eeq
is the Green's function.\footnote{If the Green's function is defined as
usual by
\beq
(\del_t -\frac{1}{2}\del^2_x){\cal G}(x, t)=\delta(x)\delta(t), \nonumber
\eeq
 \beq
{\cal G}(x,t)= \theta(t)G(x,t),\nonumber
\eeq
where $\theta(t)$ is the step function.}
We expand $u$ as
\beq
u(x, t)=u_0(x, t) + \eps u_1(x, t) + \cdots .
\eeq
As an initial condition, we take
\beq
u(x, 0)=u_0(x, 0)=\frac{m_0}{\sqrt{2\pi l^2}}{\rm e}^{-\frac{x^2}{2l^2}},
\eeq
where $m_0$ and $l$ are parameters.
Then one finds\cite{goldenfeld1} that
\beq
u(x, t)= \frac{m_0}{\sqrt{2\pi t}}{\rm e}^{-\frac{x^2}{2t}}
\{1 - \frac{\eps}{\sqrt{2\pi e}}\ln \frac t{l^2}\} + O(\eps ^2)
 + O(l^2/t, \eps),
\eeq
where we have retained in $u_1$ only the term which may contribute to the
 anomalous dimension.

We renormalize $m_0$ at $t=t_0$ as follows;
\beq
m(t_0)= Z(t_0, l)m_0, \ \ \ Z(t_0, l)=
1 - \frac{\eps}{\sqrt{2\pi e}}\ln \frac {t_0}{l^2},
\eeq
accordingly,
\beq
u(0, t_0)= m(t_0)/\sqrt{2\pi t_0},
\eeq
 and then
\beq
u(x, t; t_0)= \frac{m}{\sqrt{2\pi t}}{\rm e}^{-\frac{x^2}{2t}}
\{1 - \frac{\eps}{\sqrt{2\pi e}}\ln \frac {t}{t_0}\} + O(\eps ^2)
 + O(l^2/t, \eps).
\eeq
Here we have made the $t_0$ dependence of $u$ explicit.

We now leave the line of argument of Goldenfeld et al\cite{goldenfeld1}.
We have a family of surfaces represented by $u(x, t; t_0)$;  the surfaces
 are parametrized by $t_0$.
Let us obtain the envelope $E$ of this family of surfaces.  We can suppose
that  the tangent curve $C_{t_0}$ (the characteristic curve) is along
$t=t_0$:
\beq
t=t_0,\ \ \ u=u(x, t_0;t_0).
\eeq
Now according to the theory given in Appendix A,
the envelope is given by the following condition;
\beq
\frac{\del u}{\del t_0}= 0 \ \ \ {\rm with}\ \ \ t_0=t.
\eeq
 This equation is reduced to  an equation for $m$,
\beq
\dot{m}+ \alpha \frac{m}{t}=0,\ \ \
\alpha = \eps/\sqrt{2\pi e},
\eeq
 hence
\beq
m(t)=\bar{m}t^{-\alpha}
\eeq
with  $\bar m$ being a constant number.  Thus the envelope is given by
\beq
u_{_E}(x, t)=u(x, t ;t)= \bar{m}t^{-(1/2 + \alpha)}F(x^2/t),
\eeq
hence the anomalous exponent reads $\alpha = \eps/\sqrt{2\pi e}$, which
of course coincides with the result of Goldenfeld et al.\footnote{It
should be mentioned that $u_{_E}(x, t)$ thus obtained does {\em  not}
 satisfy  the Barlenblatt's equation even in the order of $\eps$;
 this is because the Gaussian form for $F(x^2/ t)$ is not correct.
 To obtain the precise form of $F(x^2/t)$, one may insert $u_{_E}(x, t)$
into the Barlenblatt's equation and solve $F(\xi)$.}

\newpage
\newcommand{\NG}{N. \ Goldenfeld}
\newcommand{\YO}{Y.\ Oono}

\end{document}